\documentclass[reprint,showkeys,showpacs,floatfix,longbibliography,amsmath,amssymb]{revtex4-2}
\usepackage{physics}
\usepackage{graphicx}
\usepackage{amsfonts}
\usepackage{epsfig}
\usepackage{fancyhdr}
\usepackage[titletoc]{appendix}
\usepackage[english]{babel}
\usepackage[colorlinks=true]{hyperref}
\usepackage[capitalize]{cleveref}
\usepackage{xcolor}
\usepackage{setspace}
\usepackage{natbib}
\usepackage{subcaption}
\begin{document}
\title{Switching global correlations on and off in a many-body quantum state by tuning local entanglement}
\author{Colin Benjamin}
\email{colin.nano@gmail.com}
\author{Aditya Dash}
\affiliation{School of Physical Sciences, National Institute of Science Education and Research, HBNI, Jatni-752050, India}
\affiliation{Homi Bhabha National Institute, Training School Complex, Anushaktinagar, Mumbai
400094, India}
\begin{abstract}
A quantum many-body state built on a classical 1D Ising model with locally entangled qubits is considered. This setup can model an infinite-player quantum Prisoner's dilemma game with each site representing two entangled players (or qubits). The local entanglement $\gamma$ between two qubits placed on a site in the 1D Ising model and classical coupling between adjacent sites of the Ising model has an apposite influence on qubits. It points to a counter-intuitive situation wherein local entanglement at a site can exactly cancel global correlations, signaling an artificial quantum many-body state wherein, by locally tuning the entanglement at a particular site, one can transition from a strongly correlated quantum state to an uncorrelated quantum state and then to a correlated classical state. In other words, we can simulate a state similar to a Type II superconducting state via local tuning of entanglement in a 1D Ising chain with entangled qubits.
\end{abstract}
\maketitle

\textbf{This work checks whether global correlations in a quantum many-body state can be tuned via local entanglement. It is crucial as it can be the stepping stone to design an artificial Type II superconductor. Our work demonstrates that a quantum many-body state can be brought to fruition using a classical 1D Ising model with locally entangled qubits as its basis. The local entanglement between two qubits placed on an Ising spin site, along with the classical coupling between adjacent sites, can influence the quantum many-body state in such a way as to cancel global correlations exactly. It signals an artificial quantum many-body state wherein, by locally tuning the entanglement at a particular site, one can transition from a strongly correlated quantum state to an uncorrelated quantum state and then to a correlated classical state. In other words, we can simulate a state like a Type II superconductor via local tuning of entanglement in a 1D Ising chain with entangled qubits. This paper establishes a substantial connection between quantum game theory and statistical mechanics with quantum many-body information. It Introduces a new technique to map a 1D Ising model chain with entangled qubits at each site to a type II superconducting state. Quantum many-body state and type II superconductors are a timely topic of high interest, and our work will impact the multidisciplinary quantum information community.}
\section{Introduction}

In this paper, we look at the correlation function of the Ising model in the context of the infinite player generalization of quantum prisoner's dilemma (QPD) and study the interplay of classical coupling between spin sites and the quantum entanglement between players/qubits at each site. Our main result is that we encounter a situation wherein global correlations vanish at a critical value of local entanglement $\gamma_0$. Further, we come across two second-order phase transitions in the zero temperature limit. The first at $\gamma_1$ indicates a phase transition from a classical phase to a random phase, and the second at $\gamma_2$ indicates a phase transition from a random phase to a quantum phase. It points to a quantum many-body state resembling a Type II superconducting state.

This paper is set up as follows. First, we introduce the 1D Ising model with a detailed derivation of the correlation function and briefly overview the Nash equilibrium mapping method between the 1D Ising model and a general symmetric game\cite{ben-epjb}. Next, we introduce the QPD game and define the correlation function regarding the game parameters. Unlike the classical prisoner's dilemma(CPD), QPD is played with qubits and local unitaries: quantum or classical replace strategies. The qubits could be horizontally or vertically polarized photons or electrons with spins up or down\cite{Eisert1999}.

We show that the strategies of quantum and cooperate are equivalent in the game. For the strategies of quantum versus classical (defect), we observe a phase transition between quantum and classical (defect) strategies at critical entanglement $\gamma_0$. We show that at the transition point, classical correlations vanish. We also analyze second-order phase transitions in the zero temperature limit at entanglement parameter values $\gamma_1$ and $\gamma_2$. $\gamma_1$ marks the transition from a classical phase to a random phase while $\gamma_2$ represents a transition from a random to a quantum phase. The striking aspect of our work is an analogy with a type-II superconductor, which we delve into at the end of the paper.

\section{1D Ising Model and mapping to a bi-matrix game}
The emergence of cooperation in one-shot social dilemmas wherein the Nash equilibrium is to defect is an ongoing field of research\cite{ben-epjb}. One such social dilemma is the game of prisoner's dilemma, where players are better off choosing defects in order to avoid incurring losses. For a brief introduction to the prisoner dilemma, see~\cite{Devos2016}. A solution via quantum mechanics was proposed in~\cite{Eisert1999,bordg}; it involves quantizing the prisoner's dilemma and showed that quantum strategy ($Q$) becomes the Nash Equilibrium~\cite{Eisert1999,bordg}.
However, this dilemma regarding the emergence of cooperation is not just at the level of two players. To model an infinite number of players, one can turn to the spin-1/2 Ising model, which has been exhaustively studied in statistical mechanics~\cite{Baxter1982}.  In a different context, see ~\cite{kanghao}, the Ising model has recently been used to study network games. The correspondence between the Ising model and a $2$-player, $2$-strategy game has been studied before~\cite{Galam2010, Sarkar2019} by utilizing the magnetization $m$, which is the difference between the number of up and down spins in the chain. It translates in the infinite player limit of the game as a difference in the fraction of players selecting cooperate versus defect strategies and has been studied for the case of both CPD and QPD~\cite{Sarkar2019}. In QPD, quantum and classical (cooperate) are equivalent, while one sees a phase transition between quantum and classical (defect).
In infinite player QPD, the entanglement is local and occurs at a particular site of the Ising model; each site contains two locally entangled players who play a $2$-player QPD.
\subsection{1-D Ising model and Correlation function}
The 1D spin-1/2 Ising Model consists of equidistant spins on a line, with each spin being allowed to take values of $\pm 1$ \cite{Baxter1982}. The Hamiltonian $\mathcal {H}$ of Ising model
\begin{equation}
\mathcal{H} = -\mathrm{\alpha}\sum_{i=1}^N {\tau}_i {\tau}_{i+1} - \mathrm{\beta} \sum_{i = 1}^N {\tau}_i,
\end{equation}
where $\mathrm{\alpha}$ is nearest neighbor interaction, $\mathrm{\beta}$ is the external magnetic field applied, and ${\tau}_i$ is spin at site $i$.
The partition function $\mathcal{Z}$ for the Ising model is
\begin{equation}
\mathcal{Z} = \sum_{{\tau}_1,{\tau}_2,..{\tau}_N} e^{\mathrm{\zeta} (\mathrm{\alpha} \sum_{i=1}^{N} {\tau}_i {\tau}_{i+1} + \mathrm{\beta} \sum_{i=1}^{N} ({\tau}_i+{\tau}_{i+1})/2)},
\end{equation}
with $\mathrm{\zeta} = \frac{1}{\mathrm{k}_B T}$ while $\mathrm{k}_B$ is Boltzmann constant and $T$ is the temperature.
We use the transfer matrix method to derive an expression for the correlation function of a 1D Ising model~\cite{Baxter1982}. Transfer matrix $\mathcal{\nu}$ has its elements given as,
\begin{equation}
\mathcal{\nu}({\tau}_i,{\tau}_{i+1}) = e^{\mathrm{\zeta}(\mathrm{\alpha} {\tau}_i {\tau}_{i+1} + \frac{\mathrm{\beta}}{2} ({\tau}_i + {\tau}_{i+1}))}.
\end{equation}
The transfer matrix $\mathcal{\nu}$ for a $2-$spin equivalent system is,
\begin{equation}
\mathcal{\nu} = \begin{bmatrix}
\mathcal{\nu}(1,1) & \mathcal{\nu}(1,-1) \\
\mathcal{\nu}(-1,1) & \mathcal{\nu}(-1,-1) \\
\end{bmatrix} = \begin{bmatrix}
e^{\mathrm{\zeta} (\mathrm{\alpha} + \mathrm{\beta})} & e^{\mathrm{-\zeta \alpha}} \\
e^{\mathrm{-\zeta \alpha}} & e^{\mathrm{\zeta} (\mathrm{\alpha} -\mathrm{ \beta})}\\
\end{bmatrix}, \label{transfermatrix}
\end{equation}
The above transfer matrix obeys\cite{Landi}:
\begin{align}
\sum_{{\tau}_2} \mathcal{\nu}({\tau}_1,{\tau}_2) \mathcal{\nu}({\tau}_2,{\tau}_3) &= \mathcal{\nu}^2({\sigma}_1,{\sigma}_3), \label{transprop1}\\
\text{and, }\;\;\;\;\;\;\;\; {\tau}_n \mathcal{\nu}({\tau}_n,{\tau}_{n+1}) &= \mathbf{\sigma_z} \mathcal{\nu}({\tau}_n,{\tau}_{n+1}), \label{transprop2}
\end{align}
where $\mathbf{\sigma_z}$ is the Pauli matrix. The eigenvalues of transfer matrix $\mathcal{\nu}$ in \cref{transfermatrix}, $\chi_{\pm}$ are
\begin{equation}
\chi_{\pm} = e^{\mathrm{\zeta \alpha}} \left(\cosh(\mathrm \zeta \mathrm{\beta}) \pm \sqrt{\sinh^2(\mathrm{\zeta} \mathrm{\beta}) + e^{-4 \mathrm{\zeta \alpha}}}\right), \label{lambdaeigen}
\end{equation}
where $\chi_+ > \chi_-$.
Eigenvectors of the transfer matrix $\mathcal{\nu}$ \cref{transfermatrix} written based on Pauli matrices are
\begin{equation}
\ket{\mathrm{v}_+} = \begin{pmatrix}
\cos(\frac{\theta}{2}) \\
\sin(\frac{\theta}{2})
\end{pmatrix}\;\; \textnormal{ and }\;\;
\ket{\mathrm{v}_-} = \begin{pmatrix}
-\sin(\frac{\theta}{2}) \\
\cos(\frac{\theta}{2})
\end{pmatrix},
\end{equation}
with $\tan\theta = \frac{e^{-2 \mathrm{\zeta \alpha}}}{\sinh(\mathrm{\zeta} \mathrm{\beta})}$.
The partition function can be defined via the transfer matrix as
\begin{equation}
\mathcal{Z} = \sum_{{\tau}_1,...,{\tau}_N} \prod_{i=1}^{N} \mathcal{\nu}({\tau}_i,{\tau}_{i+1}).
\end{equation}
Using the property in \cref{transprop1}, the partition function simplifies to,
\begin{equation}
\mathcal{Z} = \sum_{{\tau}_1} \mathcal{\nu}^N({\tau}_1,{\tau}_1) = \Tr(\mathcal{\nu}^N),
\end{equation}
which can be further simplified as
\begin{equation}
\mathcal{Z} = \lambda_+^N + \lambda_-^N = \lambda_+^N \left( 1 + \left(\frac{\lambda_-}{\lambda_+}\right)^N\right). \label{partition_eigenvalues}
\end{equation}
In the thermodynamic limit ($N \to \infty$), second term in \cref{partition_eigenvalues} vanishes and $Z$ reduces to
\begin{equation}
\mathcal{Z} = \lambda_+^N .\label{partition_func}
\end{equation}
The free energy per spin of the Ising chain is then given as,
\begin{equation}
\mathcal{F} = - \frac{1}{\mathrm{\zeta} N} \ln(\mathcal{Z}) = \frac{1}{\mathrm{\zeta}} \ln(\lambda_+) .
\end{equation}
The magnetization $\mathrm{m}$ is then derived using \cref{lambdaeigen} as
\begin{equation}
\mathrm{m} = -\pdv{\mathcal{F}}{\mathrm{\beta}} = \frac{\sinh(\mathrm{\zeta}\mathrm{\beta})}{\sqrt{\sinh^2(\mathrm{\zeta} \mathrm{\beta}) + e^{-4 \mathrm{\zeta \alpha}}}}. \label{isingmag}
\end{equation}
Correlation function $G(q)$~\cite{Baxter1982} is defined as,
\begin{equation}
G(q) = \frac{1}{\mathcal{Z}} \sum_{{\tau}_1,...,{\tau}_N} {\tau}_n\; {\tau}_{n+q}\; \prod_{i=1}^{N} \mathcal{\nu}({\tau}_i,{\tau}_{i+1}) ,\label{Correlation_def}
\end{equation}
where $q$ is the distance between two spin sites.
Utilizing \cref{transprop1,transprop2} in \cref{Correlation_def} and summing over all spins, we get,
\begin{equation}
G(q) = \frac{1}{\mathcal{Z}} \Tr(\mathcal{\nu}^{n-1} \mathbf{\sigma_z} \mathcal{\nu}^{q}\mathbf{\sigma_z} \mathcal{\nu}^{N-n-q+1}).
\end{equation}
Computing trace in eigenvalue basis and taking the thermodynamic limit\cite{Landi}, we obtain the final form of the correlation function as
\begin{equation}
G(q) = \cos^2(\theta) + \left(\frac{\chi_-}{\chi_+}\right)^q \sin^2(\theta) \label{correlationfunction}.
\end{equation}
In terms of the Ising model parameters, the correlation function reads as
\small{
\begin{equation}
G(q) = \frac{e^{4 \mathrm{\zeta \alpha}} + \csch^2 (\mathrm{\zeta} \mathrm{\beta}) \left(\dfrac{\cosh (\mathrm{\zeta} \mathrm{\beta}) - \sqrt{e^{ -4\mathrm{\zeta \alpha}} + \sinh^2(\mathrm{\zeta}\mathrm{ \beta})}}{\cosh (\mathrm{\zeta} \mathrm{\beta}) + \sqrt{e^{-4 \mathrm{\zeta \alpha} } + \sinh^2(\mathrm{\zeta} \mathrm{\beta})}}\right)^q} {e^{4 \mathrm{\zeta \alpha}} + \csch^2 (\mathrm{\zeta} \mathrm{\beta})}. \label{isingcor}
\end{equation}
}
\normalsize
A positive value of the correlation function indicates the probability that two spins at two sites separated by a distance $q$ will have exact alignment. A negative value indicates the probability that two spins at two sites will have opposite values.
In the limit of $T \to 0$, we have $G(q) \to 1$ for $\mathrm{\alpha} > 0$ as all spins align in the same direction, while for $\mathrm{\alpha} < 0$, we find the correlation function to be $G(q) \to (-1)^q$ as the nearest neighboring spins are oppositely aligned. At $T \to \infty$, $G(q) \to 0$ as all spins are entirely randomized for all $\mathrm{\alpha}$.

\subsection{Mapping between 1D Ising model and a bi-matrix game \label{mapping-game-to-ising}}
To examine a game with infinite players, we map a $2$-player, $2$-strategy game to the Ising model and consider the thermodynamic limit. A recipe for such a mapping is given in \cite{Galam2010}. We here outline the important aspects of successful mapping. To make a correspondence between the 1D Ising model and a general bi-matrix game, first, we consider a $2$-site 1D Ising Hamiltonian, which is given by
\begin{equation}
\mathcal{H} = \mathrm{-\alpha} \tau_2 {\tau}_1 -\mathrm{\alpha} {\tau}_1 {\tau}_2 - \mathrm{\beta} ({\tau}_1 + {\tau}_2) = \mathcal{E}_1 + \mathcal{E}_2
\end{equation}
${\tau}_1, {\tau}_2$ are two spins in the Ising Hamiltonian, and $\mathrm{\alpha}$ is the coupling between two spins while $\mathrm{\beta}$ is the external magnetic field. The Hamiltonian can be split into two parts, which correspond to the energy of two sites and spins as
\begin{equation}
\mathcal{E}_1 = \mathrm{-\alpha} {\tau}_1 {\tau}_2 - \mathrm{\beta} {\tau}_1 \;\;\;\;\textnormal{and}\;\;\;\; \mathcal{E}_2 = \mathrm{-\alpha} {\tau}_2 {\tau}_1 - \mathrm{\beta} {\tau}_2
\end{equation}
The energy for the first site/spin can be written in matrix form as
\begin{equation}
\mathcal{E} = \left(\begin{array}{c|cc}
& +1 & -1 \\ \hline
+1 & \mathrm{\alpha}+\mathrm{\beta} & -(\mathrm{\alpha}-\mathrm{\beta}), \\
-1 & -(\mathrm{\alpha}+\mathrm{\beta}) & \mathrm{\alpha}-\mathrm{\beta} \\
\end{array}\right). \label{isingmatrix}
\end{equation}
In the Ising Hamiltonian, we minimize energy to obtain the equilibrium solution. On the other hand, in the case of a game, players try to reach the Nash equilibrium, where players maximize their payoffs. Therefore, to correspond with the game, we take the negative of energy such that minimizing \cref{isingmatrix} implies maximizing the system's energy, leading to a Nash Equilibrium.

A general $2$-player, $2$-strategy game is given by the payoff matrix
\begin{equation} \label{gendoublepayoffmatrix}
\mathcal{P}=\left(\begin{array}{c|cc}
& st_1 & st_2 \\\hline
st_1 & x,x' & y,y' \\
st_2 & z,z' & w,w'
\end{array}\right),
\end{equation}
where $\mathcal{st}_1, \mathcal{st}_2$ are the strategies available to players and $x,y,z,w$ are the payoffs associated with row player and $x',y',z',w'$ are the payoffs for column player. If the game is symmetric, then $x = x', y= z', z = y'$ and $w = w'$. The payoff matrix for a symmetric matrix is only written for row players (See, \cite{Devos2016} for a brief explanation).
In such a case, the general symmetric game can be described by the payoff matrix as
\begin{equation}
\mathcal{P = \left(\begin{array}{c|cc}
& st_1 & st_2 \\ \hline
st_1 & x & y \\
st_2 & z & w \\
\end{array}\right).} \label{genpayoff}
\end{equation}
To map the payoff matrix to the Ising model, we transform the first column by adding $\lambda = -(x+z)/2$ and the second column by adding $\mu = -(y+w)/2$ \cite{Galam2010}. This results in the payoffs having the same structure as the energies of the two-spin Ising model; see~\cref{isingmatrix}. This transformation does not change the Nash Equilibrium of the game (see \cite{Galam2010, Devos2016}) such that the payoff matrix \cref{genpayoff} is symmetrical to \cref{isingmatrix},
\begin{equation}
\mathcal{P = \left(\begin{array}{c|cc}
& st_1 & st_2 \\ \hline
st_1 & \frac{x-z}{2} & \frac{y-w}{2} \\
st_2 & \frac{z-x}{2} & \frac{w-y}{2} \\
\end{array}\right). }\label{genpayoff-trans}
\end{equation}
The proof regarding the statement above that the transformation does not change the Nash equilibrium is explicitly shown in Appendix of Ref.~\cite{Sarkar2018b}) using Brouwer's fixed point theorem.
The $\mathrm{C}$ and $\mathrm{B}$ factors for the game can be expressed in terms of payoffs of the game by comparing \cref{genpayoff-trans} and \cref{isingmatrix},
\begin{equation}
\mathrm{\alpha} = \frac{x-y+w-z}{4}, \;\;\;\; \textnormal{and} \;\;\;\ \mathrm{\beta} = \frac{x-z+y-w}{4}. \label{genrelation}
\end{equation}

\section{Infinite player quantum Prisoner's dilemma as a quantum many-body state}
\begin{figure}[htbp]
\centering
\includegraphics[width=\linewidth]{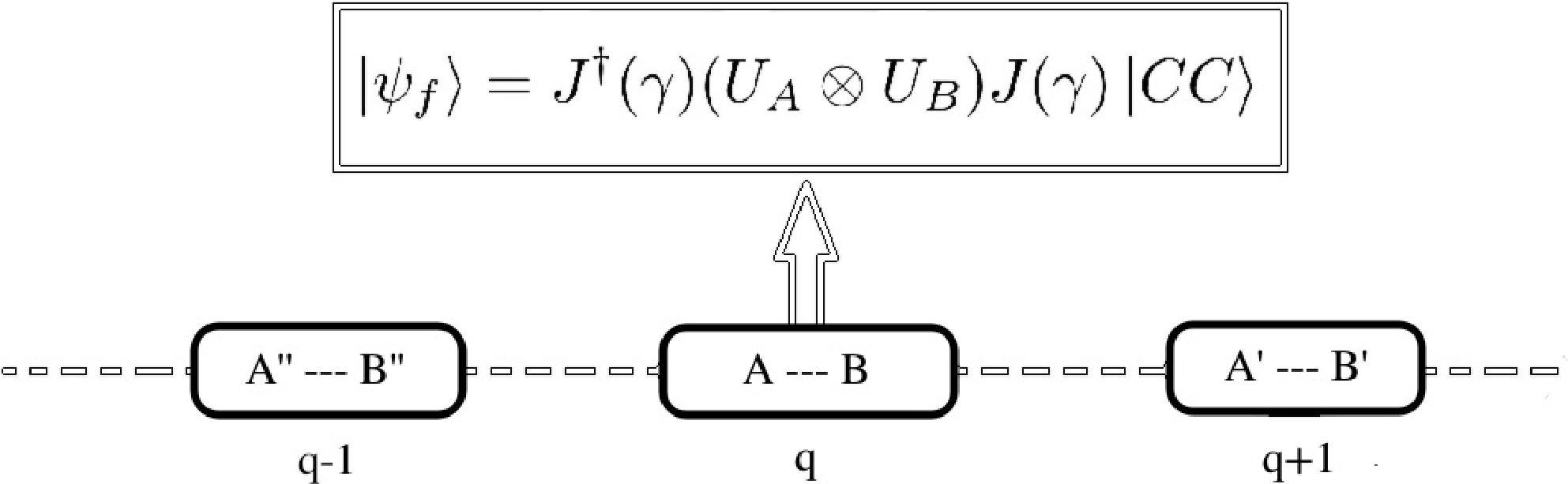}
\caption{Implementation of QPD game in the thermodynamic limit. Two players/qubits are situated at each site of the Ising lattice (A and B, for example, at site $q$) and are entangled with each other while the sites, in turn, are classically coupled. The external field here influences all sites equivalently to select identical strategies.}
\label{setup}
\end{figure}

A QPD is the quantized version of a CPD obtained using the scheme of Ref.~\cite{Eisert1999,bordg}. In QPD, players are represented by qubits (say, A and B), which are in basis $\ket{0}$ and $\ket{1}$, representing the cooperate and defect states, respectively. The qubits associated can be entangled via an entangling operator $\hat{J}(\gamma) = \cos \left(\dfrac{\gamma}{2} \right) I \otimes I - i \sin \left(\dfrac{\gamma}{2} \right) Y \otimes Y$ at the beginning of game. Here, $Y = i \sigma_y$ and $ \gamma \in [0,\pi/2]$
with $\gamma$ being the degree of entanglement.
The general form of the unitary operator which acts on the qubits is,
\begin{equation}
U(\delta,\Delta) = \begin{bmatrix}
e^{i \Delta}cos(\delta /2) & sin(\delta /2) \\
- sin(\delta /2) & e^{-i \Delta}cos(\delta /2) \\
\end{bmatrix}.
\end{equation}
The classical strategies of cooperate and defect are given by $I$ and $X$ ($= \sigma_x$), respectively. The game is initialized with both qubits in state $\ket{0}$. The qubits are then entangled via $\hat{J}(\gamma)$. Operators(or strategies in game theory parlance) are then applied to the entangled qubits. Finally, the qubits are disentangled by a disentangling operator $\hat{J}^\dagger(\gamma)$ to obtain final state $\ket{\psi_f}$, where $\ket{\psi_f}$ is given by
\begin{equation}
\ket{\psi_f} = \hat{J}^\dagger(\gamma) (U_A \otimes U_B) \hat{J}(\gamma) \ket{00}
\end{equation}
The payoffs for either qubit are then given as,
\begin{align}
P_A &= x P_{CC} + w P_{DD} + z P_{DC} + y P_{CD}, \\ \textnormal{and,} \;\;\;
P_B &= x P_{CC} + w P_{DD} + y P_{DC} + z P_{CD}.
\end{align}
with $P_{GH} = |\braket{GH}{\psi_f}|^2$, and ${G, H}\in \{C, D\}$ and $x, y, z, w$ being the payoffs colloquially ascribed to reward, suckers payoff, temptation, and punishment respectively in CPD\cite{Devos2016}.
The quantization scheme introduces a new strategy $Q = I \sigma_z$, the quantum strategy(or operator) for both qubits. The payoff matrix becomes,
\begin{equation}
P = \left(\begin{array}{c|ccc}
& C & D & Q \\ \hline
C & x & y & y1 \\
D & z & w & y2 \\
Q & y1 & y3 & x \\
\end{array}\right)
\end{equation}
where $y1 = x \cos^2(\gamma) + w \sin^2(\gamma), \; y2 = z \cos^2 (\gamma) + y \sin^2 (\gamma)$ and $y3 = y \cos^2 (\gamma) + z \sin^2 (\gamma)$.
Now, for the infinite player/qubit version of QPD, we arrange the qubits using the setup depicted in\cref{setup}. The setup is such that entanglement is local at a particular site that contains two qubits that play a two-player/qubit QPD, while the coupling between two such adjacent sites is classical.
Since the payoff matrix is a $3\times3$ matrix, while the Ising matrix is a $2\times2$ matrix, we consider the game in two cases: quantum versus cooperate and quantum versus defect.

\subsection{Quantum versus Cooperate}
In this case, the quantum (Q) or the classical cooperate strategy (C) is used by the players/qubits. The payoff/output matrix for the quantum versus cooperate case is,
\begin{equation}
P = \left(\begin{array}{c|cc}
& Q & C \\ \hline
Q & x & x \cos^2(\gamma) + w \sin^2(\gamma) \\
C & x \cos^2(\gamma) + w \sin^2(\gamma) & x \\
\end{array}\right). \label{QvsC}
\end{equation}

The transformations needed to map~\cref{QvsC} to Ising model as outlined in \cref{mapping-game-to-ising} are $\lambda = -\frac{x(1+\cos^2(\gamma))+w\sin^2(\gamma)}{2}$ and $\mu = -\frac{x(1+\cos^2(\gamma))+ w \sin^2(\gamma)}{2}$. We get the equivalent $\mathrm{\alpha}$ and $\mathrm{\beta}$ factors as
\begin{equation}
\mathrm{\alpha} = \frac{(x-w)\sin^2 (\gamma)}{2} ,\;\;\;\; \textnormal{and} \;\;\;\; \mathrm{\beta} = 0. \label{QvsCrel}
\end{equation}
Since $\mathrm{h} = 0$, we find that the game magnetization
\begin{equation}
\mathrm{m_g} = \frac{\sinh(\mathrm{\zeta} \mathrm{\beta})}{\sqrt{\sinh^2(\mathrm{\zeta}\mathrm{\beta}) + e^{-4 \mathrm{\zeta \alpha}}}} = 0.
\end{equation}
It implies that the players/qubits randomly oscillate between two strategies and that both are equivalent in this game. The correlation function $G(q)$ for this case is obtained by substituting \cref{QvsCrel} in \cref{isingcor} and taking limit $\mathrm{\beta} \to 0$ in \cref{isingcor}, since the correlation function is not determinate at $\mathrm{\beta=0}$ \cite{Landi},
\begin{equation}
\label{cor-coop}
G(q) = \tanh \left(\frac{(x-w)\sin^2 (\gamma)}{2 T} \right)^q,
\end{equation}
wherein $T = (\mathrm{k_B}\mathrm{\zeta})^{-1}$ is the game temperature. We infer from~\cref{QvsCrel} that the classical coupling between sites depends upon the degree of entanglement between the two qubits at a site. For $\gamma = 0$, we have $\mathrm{\alpha} = 0$, implying that all sites are disconnected, and each player is uncorrelated in the game, hence $G(q) \to 0$. At maximal entanglement ($\gamma = \pi/2$), $\mathcal{\alpha} = 1$ for $x = 3$ and $w = 1$ which leads to very high correlation among qubits. Taking the limit $T \to 0$, we find that $G(q) \to 1$, which implies that all the qubits are wholly correlated, while for limit $T \to \infty$, $G(q) \to 0$ as all qubits in the game act randomly in changing from quantum to cooperate. We plot the game magnetization and the correlation function for two different distances $q$ as a function of the entanglement in~\cref{figure_coop}.
\subsection{Quantum versus Defect}
In this case, qubits are restricted to the quantum or classical (defect) operators. So, the payoff matrix for the quantum versus defect case becomes
\begin{equation}
P = \left(\begin{array}{c|cc}
& Q & D \\ \hline
Q & x & y \cos^2(\gamma) + z \sin^2(\gamma) \\
D & z \cos^2(\gamma) + y \sin^2(\gamma) & w \\
\end{array}\right). \label{QvsD}
\end{equation}
From the payoff matrix, we find that Nash Equilibrium for $\gamma = 0$ is the classical operation (defect), while at maximal entanglement $\gamma = \pi/2$, Nash Equilibrium is the quantum strategy.
The transformations needed to map the two-player case to the infinite player limit as per the recipe given in \cref{mapping-game-to-ising} are $\lambda = -\frac{x+z\cos^2(\gamma)+y \sin^2(\gamma)}{2}$ and $\mu = -\frac{w + z\sin^2(\gamma)+ y\cos^2(\gamma)}{2}$.
The equivalent $\alpha$ and $\beta$ factors for this case are then obtained as
\begin{equation}
\!\!\!\!\!\!\mathrm{\alpha} = \frac{x+w-z-y}{4},\!\!\!\!\;\;\;\;\mbox{ } \!\!\mathrm{\beta} = \frac{x-w + (y-z)\cos(2\gamma)}{4}. \label{qpdrel}
\end{equation}
Since the coupling $\mathrm{\alpha}$ between sites is independent of the entanglement $\gamma$, it is entirely classical in contrast to the case of quantum versus cooperate, where the strength of the coupling $\alpha$ is dependent on the degree of entanglement $\gamma$.
The external field $\mathrm{\beta}$ is a function of entanglement $\gamma$ and is non-zero, in stark contrast to the quantum versus cooperate case where the external field is absent.

The game magnetization $\mathrm{m_g}$ is
\begin{equation}
\mathrm{m_g } = \frac{\sinh \left(\frac{-w+x+\cos (2 \gamma ) (y-z)}{4 T}\right)}{\sqrt{\sinh ^2\left(\frac{-w+x+\cos (2 \gamma ) (y-z)}{4 T}\right)+e^{\frac{-w-x+y+z}{T}}}}.
\label{gamemag}
\end{equation}

The game magnetization differentiates the fraction of players/qubits playing quantum versus classical (defect) strategies. The number of players/qubits playing quantum and classical (defect) strategies are identical for zero magnetization.
For zero entanglement, $\mathrm{m_g} \to 0$ as $T \to \infty$ as players/qubits randomly alternate between quantum and defect, while for $T \to 0$, we find that $\mathrm{m_g} \to -1$, implying that all qubits have to defect.
For maximal entanglement, $\mathrm{m_g} \to 0$ as $T \to \infty$ since players/qubits randomly change from quantum to defect while at $T \to 0$, we find that $\mathrm{m_g} \to 1$, implying that all players/qubits have to opt for quantum.

Correlation function $G(q)$ for the case of quantum versus defect is obtained by substituting~\cref{qpdrel} in\cref{correlationfunction} and is given as
\begin{widetext}
\begin{equation}
\label{gamecor}
G(q) = \frac{e^{\frac{y+z}{T}} \text{csch}^2\left(\frac{-w+x+\cos (2 \gamma ) (y-z)}{4 T}\right) \left(\frac{\cosh \left(\frac{-w+x+\cos (2 \gamma ) (y-z)}{4 T}\right)-\sqrt{\sinh ^2\left(\frac{-w+x+\cos (2 \gamma ) (y-z)}{4 T}\right)+e^{\frac{-w-x+y+z}{T}}}}{\sqrt{\sinh ^2\left(\frac{-w+x+\cos (2 \gamma ) (y-z)}{4 T}\right)+e^{\frac{-w-x+y+z}{T}}}+\cosh \left(\frac{-w+x+\cos (2 \gamma ) (y-z)}{4 T}\right)}\right)^q+e^{\frac{w+x}{T}}}{e^{\frac{y+z}{T}} \text{csch}^2\left(\frac{-w+x+\cos (2 \gamma ) (y-z)}{4 T}\right)+e^{\frac{w+x}{T}}}
\end{equation}
\end{widetext}
A positive correlation between two sites at a distance $q$ implies that they will behave similarly, while a negative correlation implies that two sites will behave oppositely. If the correlation is zero, the two sites do not influence each other. As explained, each site has $2$ players/qubits with the QPD algorithm being played (See~\cref{setup}).

\section{Many body quantum correlations }

\subsection{At finite temperatures. \label{finiteT}}

\begin{figure*}[htbp]
\centering
\includegraphics[width = \linewidth]{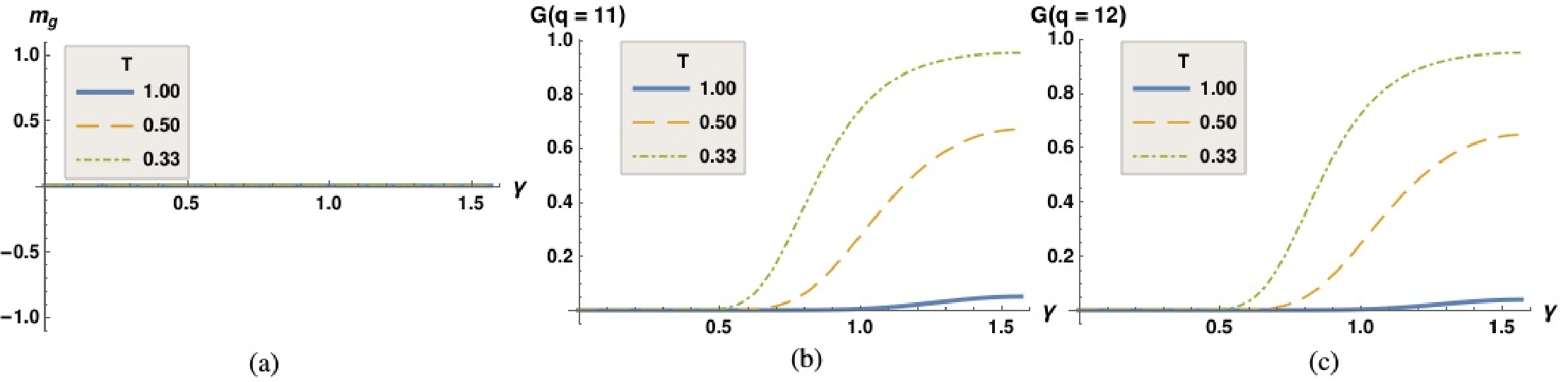}
\caption{Plot of (a) game magnetization $\mathrm{m_g}$ and (b) correlation function $G(q)$ for $q = 11$ and (c) correlation function $G(q)$ for $q = 12$ versus entanglement $\gamma$ for quantum vs cooperate strategies. Here, $x =3, w=1,y=0$ and $z = 5$.}
\label{figure_coop}
\end{figure*}

\begin{figure*}[htbp]
\centering
\includegraphics[width = \linewidth]{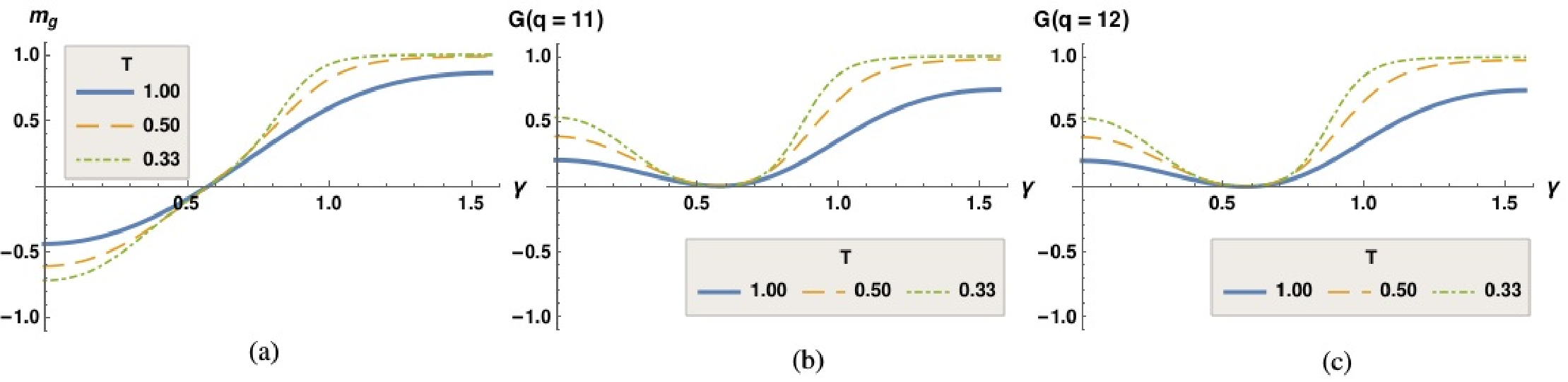}
\caption{Plot of (a) game magnetization $\mathrm{m_g}$ and (b) correlation function $G(q)$ for $q = 11$ and (c) correlation function $G(q)$ for $q = 12$ versus entanglement $\gamma$ for quantum versus defect case. Here, $x =3, w=1,y=0$ and $z = 5$}
\label{figure}
\end{figure*}

\subsubsection{Quantum versus Cooperate}
We plot the correlation function $G(q)$\cref{gamecor} for the quantum many-body state in the case of quantum versus cooperate in~\cref{figure_coop} and quantum versus defect in~\cref{figure}. In the quantum versus cooperate setup of QPD, the players/qubits at a particular site do not distinguish between strategies of quantum and cooperate as game magnetization is zero for all entanglement values as shown in~\cref{figure_coop}(a). The correlation function shows that, beyond a particular value, the entanglement between two players/qubits at a site improves the classical correlation between two sites situated far apart. It results from coupling $J$ being positive, which makes the sites behave similarly.

\subsubsection{Quantum versus Defect}
In the quantum versus defect case, from the plot of game magnetization \cref{figure}(a), we observe that the qubits switch their preferred strategy from defect to quantum as the magnetization changes sign from positive to negative. The transition point occurs when $\mathrm{m_g = 0}$. From~\cref{gamemag}, the transition point is at
\begin{equation}
\gamma_0 = \frac{1}{2}\arccos(\frac{x-w}{z-y}). \label{gamma0}
\end{equation}
$\gamma_0 \approx 0.579$ for $x =3.0, w= 1.0, y=0.0$ and $z=5.0$ in accordance with payoffs for prisoner's dilemma.

We observe that at $\gamma_0$, the correlation between sites becomes zero. It can be attributed to the local entanglement between qubits at a site negating the classical correlation between sites of the infinite player/qubit game. It results in all sites behaving independently, which leads to qubits opting for either quantum or defect strategy with equal probability. It implies that the coupling $\alpha$ between qubits and the local entanglement $\gamma$ between qubits at a particular site have competing influences. The coupling forces the sites to choose defection, while entanglement makes qubits choose quantum.
For entanglement values below the transition point, classical coupling between sites dominates the local entanglement, while above the transition point, entanglement is dominant. Also, from these plots, we can determine that the entanglement at a site is much stronger than the coupling between sites as qubits adopt cooperation at maximal entanglement.

Below critical entanglement, players/qubits are correlated as they choose to defect, while above the critical value, qubits are correlated as they choose to select quantum strategy. It is important to note here that the presence of quantum entanglement is responsible for vanishing classical correlations, unlike the case in~\cite{Kaszlikowski2008}, where there is no direct evidence to suggest that the classical correlations are killed off by quantum entanglement. Also, the entanglement in this game is bipartite as it occurs between two qubits at a site.

\subsection{ $T \to \infty$ and $T \to 0$ limit}
\begin{figure*}[htbp]
\centering
\includegraphics[width = \linewidth]{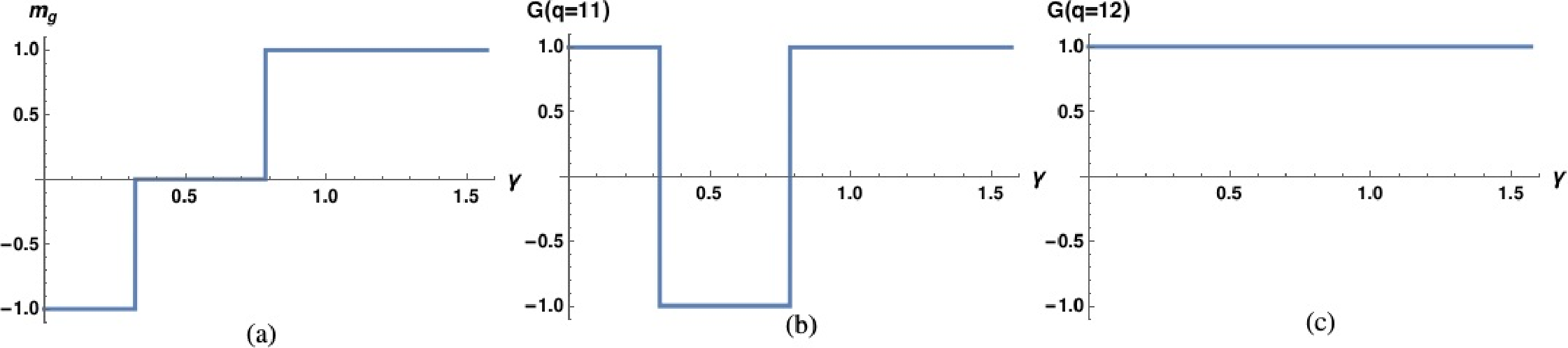}
\caption{Plot of (a) game magnetization $\mathrm{m_g}$ and (b) correlation function $G(q)$ for $q = 11$ and (c) correlation function $G(q)$ for $q = 12$ versus entanglement $\gamma$ in the $T \to 0$ limit. Here, $x =3,w=1,y=0$ and $z = 5$}
\label{T0fig}
\end{figure*}

In \cref{finiteT}, we have analyzed the correlation function for quantum versus defect case at finite $T$ and found a critical point $\gamma_0$ where classical correlations vanish.
Taking limit $T \to \infty$, we find that the correlation function $G(q) \to 0$ for all distances $q$ and all entanglement $\gamma$ as qubits are operated on by either unitary with complete randomness. At the $T \to 0$ limit, we observe three different phases in-game as plotted in~\cref{T0fig}(a), namely classical phase which exists from $0 <\gamma < \gamma_1$ where $\mathrm{m_g = -1}$, random phase existing between $\gamma_1 < \gamma < \gamma_2$ where $\mathrm{m_g = 0}$, and finally the quantum phase, which exists between $\gamma_2 < \gamma < \pi/2$ and $\mathrm{m_g = 1}$. Magnetization transitions from $-1$ to $0$ as entanglement crosses $\gamma_1$ and from $0$ to $+1$ as entanglement crosses $\gamma_2$. An analytic expression can be derived for $\gamma_1$ and $\gamma_2$ by observing that magnetization is zero if $e^{\frac{-w-x+y+z}{T}} > \sinh^2 \left(\frac{-w+x+\cos (2 \gamma ) (y-z)}{4 T}\right)$. Thus, solving
\begin{equation}
e^{\frac{-w-x+y+z}{T}} = \sinh^2 \left(\dfrac{-w+x+\cos(2\gamma)(y-z)}{4 T}\right),
\end{equation}
for $\gamma$, we get ($\gamma_1 < \gamma_2$),
\begin{equation}
\gamma_1 = \frac{1}{2} \arccos\left(\dfrac{3w+x-2(y+z)}{y-z}\right), \label{gamma1}
\end{equation}
\begin{equation}
\mbox{and }\gamma_2 = \frac{1}{2} \arccos\left(\dfrac{3x-w+2(y-z)}{y-z}\right). \label{gamma2}
\end{equation}
for the condition $w+x<y+z$. Otherwise, only two phases remain in-game: the classical phase, which exists for $0 < \gamma < \gamma_0$, and the quantum phase, which exists for $\gamma_0 < \gamma < \pi/2$, with $\gamma_0$ defined in \cref{gamma0}. $\gamma_0$ always lies between $\gamma_1$ and $\gamma_2$ and is independent of temperature $T$.
For the plots in \cref{T0fig}, $\gamma_1 \approx 0.321$ and $\gamma_2 \approx 0.785$.
It is a second-order phase transition as the game magnetization displays a discontinuity at $\gamma_1$ and $\gamma_2$ \cite{Baxter1982}.

\subsubsection{Simulating a state similar to type-II superconductors}
Type-II superconductors display two second-order phase transitions as the applied magnetic field $H$ increases for temperatures below the critical temperature \cite{kittel_2005, tinkham_1996}. We can see that the random-quantum phase transition at $\gamma_2$ in QPD can be analogous to the Meissner-vortex phase for the type-II superconductors. In contrast, the classical-random phase transition at $\gamma_1$ is similar to the vortex-normal phase transition of superconductors, where the vortex phase consists of normal regions surrounded by superconducting areas. In a type-II superconductor, there are two phase transitions: one from the normal state to a vortex state at a critical field $H_{c1}$ where the magnetic field coexists with the superconducting phase and then to a pure superconducting state at $H_{c2}$. It is replicated in our model, too, and we see a phase transition from the classical defect state where qubits at each site adopt a defect strategy (akin to the normal metal state of a type II superconductor) at a particular value of entanglement between qubits to a random state where each qubit randomly orients its strategy such that its neither pure quantum nor pure defect akin to the vortex phase of a type II superconductor. Then, at a more considerable value of entanglement, we see a second phase transition from this random state to a pure quantum state wherein qubits universally adopt the quantum strategy akin to the superconducting state.
Also, our model has two entangled qubits at every site, while the Cooper pairs in the superconductor are weakly bound but mobile throughout the material.
The spin-spin correlation of Cooper pairs in the superconductor falls off as $\exp(- \xi)$ for significant $\xi$ distance between two pairs \cite{tinkham_1996}. From \cref{gamecor}, the correlations fall off as $1/q$ for large distances $q$ between the sites.

\section{Conclusion}

This paper presents a situation where we can observe the interplay between quantum entanglement and classical correlations in a quantum many-body state. It is done by considering the infinite player limit of QPD. The infinite-player game is constructed by mapping QPD to the spin-$1/2$ Ising model in the thermodynamic limit. The game is set up so that each site in the Ising model is occupied by two players/qubits who are locally entangled. The sites are coupled by the nearest neighbor coupling, which is entirely classical. To construct the mapping, we compare only two strategies simultaneously: quantum versus cooperate or quantum versus defect.

We find that local entanglement enhances classical correlations between sites in the case of quantum versus cooperate. On the other hand, in the quantum versus defect case, we find that local entanglement and classical coupling between adjacent sites negatively influence the players/qubits. The classical coupling forces sites to opt for defect, while the entanglement at the site makes players/qubits choose quantum. It can also be seen that the entanglement is stronger than the coupling as all players/qubits choose quantum at maximal entanglement.
The players are entirely uncorrelated at the transition point $\gamma_0$, i.e., entanglement and classical coupling cancel out each other's effects, leading to players/qubits selecting their strategies randomly.
In the $T \to 0$ limit, there can be three phases, with each phase separated by a first-order transition, with the random phase constituting players/qubits such that nearest neighbors select opposite strategies while the classical and quantum strategies dominate the other two phases.
\acknowledgments This work was supported by the grant "Josephson junctions with strained Dirac materials and their application in
quantum information processing" from the Science \& Engineering Research Board (SERB), New Delhi, Government of India, under Grant No. $CRG/2019/006258$.

\bibliography{references}
\bibliographystyle{apsrev.bst}
\end{document}